\documentclass{appolb}
\usepackage{epsfig}
\usepackage{graphicx}
\newcommand{\invfb}{fb$^{-1}$\ }
\newcommand{\degr}{$^\circ$}
\newcommand{\symbolfootnote}[1]{\begingroup%
  \def\thefootnote{\fnsymbol{footnote}}\footnote{#1}\endgroup} 

\begin{document}
\preprint{\flushright CP3-06-18}
\title{GASTOF: Ultra-fast ToF forward detector\\for exclusive processes at the LHC 
\thanks{Presented by T.Pierzcha\l a at the conference ''Physics at LHC'', 
Krak\'ow, June 2006}%
}
\author{L. Bonnet, T. Pierzcha\l a\symbolfootnote {
Contact person, email: tomasz.pierzchala@fynu.ucl.ac.be},
K. Piotrzkowski, P. Rodeghiero
\address{ Centre for Particle Physics and Phenomenology (CP3)\\ 
        Universit\'e Catholique de Louvain\\
       Chemin du Cyclotron 2\\
B-1348 Louvain-la-Neuve, Belgium \\
{\it for the FP420 collaboration }}
}
\maketitle
\begin{abstract}
GASTOF (Gas Time-of-Flight) detector is a Cherenkov detector proposed 
for very precise ($\delta t\sim$ 10--20 ps) arrival time measurements of forward 
protons at some 420 m from the central detectors of CMS and ATLAS. 
Such an excellent time resolution will allow by z-by-timing technique for 
precise measurement of the z-coordinate of the 
event vertex in exclusive production at the LHC, when two colliding protons
are scattered at very small angles. 
In the paper we present first GASTOF prototype, simulations of its performance
as well as first tests using a cosmic muon telescope.

\end{abstract}
  
\section{Introduction}
In the central exclusive processes $pp\to p\oplus X \oplus p$, 
where $\oplus $ denotes the absence of hadronic activity, the central 
system X can be produced either via $\gamma\gamma$ fusion or in 
exclusive diffractive production. Detection of the forward scattered protons
will allow for new and complementary studies at the LHC.
Already at low luminosity, the two photon dilepton production, precisely known from
QED, can serve as a calibration process \cite{dis}. 
For few \invfb of the integrated $pp$ luminosity the high energy  
photon physics opens up giving access to precision studies of quartic gauge 
couplings, anomalous W or Z pair production, and at higher luminosities, 
super-symmetric particle pair production in very clean environment \cite{gg}. 
Starting from tens of \invfb the exclusive diffractive production of the 
Higgs boson becomes important \cite{Khoze:2001xm,Cox:2005if}. 

There are three main reasons 
why the exclusive diffractive production is important especially for the 
Higgs boson studies at the LHC:
\begin{enumerate}
\item Forward scattered protons, tend to "select" state of the central system X 
to be $J_z$ = 0, C and P even ($0^{++}$).
Moreover, correlations between outgoing proton momenta are sensitive to these quantum numbers.
\item Mass resolution of about 2\% of the central system can be achieved 
from momenta measurements of the scattered protons alone, which would help 
to resolve a nearly degenerate super symmetric Higgs sector, for example 
\cite{ellis}.
\item Excellent signal to background ratio of about unity for the SM Higgs 
production and more than an order of magnitude larger for certain MSSM 
scenarios.
\end{enumerate}
The FP420 collaboration proposes \cite{FP420} to install proton detectors at 
some 420 m from the interaction point (IP) of CMS or ATLAS. Acceptance of such detectors 
matches very well energy distributions of the forward protons in the exclusive 
production of the light Higgs boson. As the cross section for diffractive production of 
the SM Higgs production is expected to be small, 
1--10 fb depending of the Higgs boson mass \cite{Cox:2005if}, 
it is imperative to measure it at high LHC luminosity
with many interactions per beam crossing (event pile-up).

The aim of a very precise time measurement using
GASTOF \footnote{Or, using a complementary QUARTIC detector
which is based on quartz radiator.}
is to determine $\emph{z}$-coordinate of event vertex
using the $\emph{z}$-by-timing technique, and consequently 
to match it with the vertex measured by central detectors.
The $\emph{z}$-by-timing technique is based on the arrival time difference
for two protons detected on both sides of the IP.
If detectors of forward protons are at distance L, and the event vertex
is displaced from the nominal IP at $\emph{z}=0$ to some $\emph{z}$ 
(width of the longitudinal distribution of the interaction point is 
expected to be of 50--70 mm) than 
\begin{equation}
\Delta t = \frac{L+\emph{z}}{c} - \frac{L-\emph{z}}{c} = 
\frac{2 \emph{z}}{c}
\label{eq:Delta_t}
\end{equation}
where $\Delta t$ is the arrival time difference of the two protons to be measured 
with GASTOF detectors, and $c$ is speed of light ($\beta\approx1$ for the 
forward protons). It follows from Eq.(\ref{eq:Delta_t}) that precision of 
$\emph{z}$ measurement is $\delta\emph{z}~=~c\delta t/\sqrt{2}$.
Hence the detector time resolutions of $\delta t$~=~20 or 10 ps would result in 4 mm or 2 
mm resolutions in $\emph{z}$, respectively. For studies of exclusive 
processes at high LHC luminosity this will be essential to reduce accidental 
coincidences due to event pile-up, where the two forward protons and the 
central system X are not coming from the same interaction.
\section{GASTOF}
GASTOF is a Cherenkov gas detector -- photons produced by high energy protons traversing gas   
medium are reflected by a mirror onto a very fast photomultiplier.
In gases, thanks to small refractive index, the Cherenkov photons are 
radiated at very small angles, and propagate at speed very close to $c$, therefore
very good time resolution is expected. The detector is simple, very robust, and light --
the multiple scattering induced by GASTOF should be small and allow 
for placing it in front of or between the planes of the proton tracking detectors, without
affecting the eventual resolutions. The FP420 forward detector sensitive areas will be 
heavily irradiated, in particular by single diffractive protons produced at huge rates at the LHC --
if needed the gas can be flushed therefore it is expected that GASTOF will be radiation hard.
In addition, thanks to its directionality and relatively
high energy threshold of incident particles GASTOF will not be too sensitive to stray 
charged particles in the LHC tunnel. On the other hand, in gas the Cherenkov photon yield is not large, 
and the radiator length is limited -- therefore, as the photon spectrum is peaked at short wavelengths, a 
lot of effort is put into providing good efficiency of detecting UV photons.

\begin{figure}[!ht]
\begin{center}
\epsfig{file=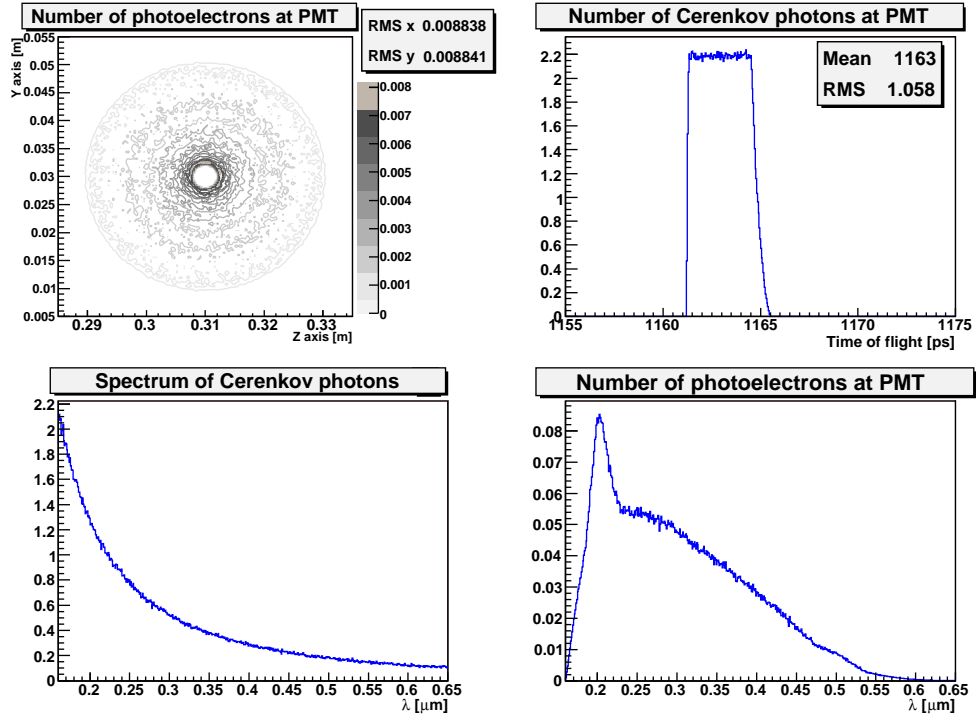,width=135mm}
\end{center}
\caption
{Results of simulation of the GASTOF prototype -- 31cm long, filled with C$_4$F$_{10}$ 
at 1.2 atm, with refractive index n$\sim$1.0018. One expects that in average, for one high energy proton
hitting GASTOF centrally, about 198 Cherenkov photons hit the mirror, and 13.4 photoelectrons are produced
at the Burle MCP-PMT. Upper, left plot shows spatial distribution of the photons at the photocathode. Upper, 
right plot shows arrival time of these photons, where $t=0$ is set for a proton entering GASTOF. Lower plots
show, respectively, numbers of produced photons and photoelectrons as a function of the photon wavelength. 
Each plot shows number of events per bin, hence sums of the bin contents are equal to the total
number of photons and photoelectrons, respectively.
}
\label{SimulationPlot}
\end{figure}
First two identical GASTOF prototypes have been built to test these expectations. A 31 cm long, 6 cm square tube, 
is filled with C$_4$F$_{10}$ at 1.2 atm, with refractive index n$\sim$1.0018. A flat mirror at
45\degr\ reflects Cherenkov light onto a 2 inch square photocathode of the micro-channel plate 
photomultiplier (MCP-PMT) 85011-501 from Burle \cite{burle}). Special UV coated mirrors have been used, 
which have non-zero reflectivity for $\lambda >$ 160 nm and more than 75\% above 180 nm. 
The Burle MCP-PMTs have UV grade fused silica windows, the collection efficiency of 50\%, and multiple anodes 
in  form of 8x8 anode matrix. They are characterized by a sharp rise time of 300 ps and low transit 
time jitter of about 40 ps.

A simple, based on ray-tracing, Monte Carlo simulation has been prepared, and its results for  
the prototype are shown in the Fig.(\ref{SimulationPlot}). For high energy charged particles hitting 
GASTOF centrally, along its axis, almost 200 photons in average are radiated and hit the mirror. This 
results in about 13 photoelectrons produced in average at the photocathode. The light spot at the photocathode
has diameter of 4 cm. Finally, all the Cherenkov photons arrive at the photocathode within a 4 ps time window!
These results indicate that indeed GASTOF can provide efficient, and extremely fast 
and accurate timing signal at the LHC.
\section{First Results}
\begin{figure}[!ht]
\begin{center}
\epsfig{file=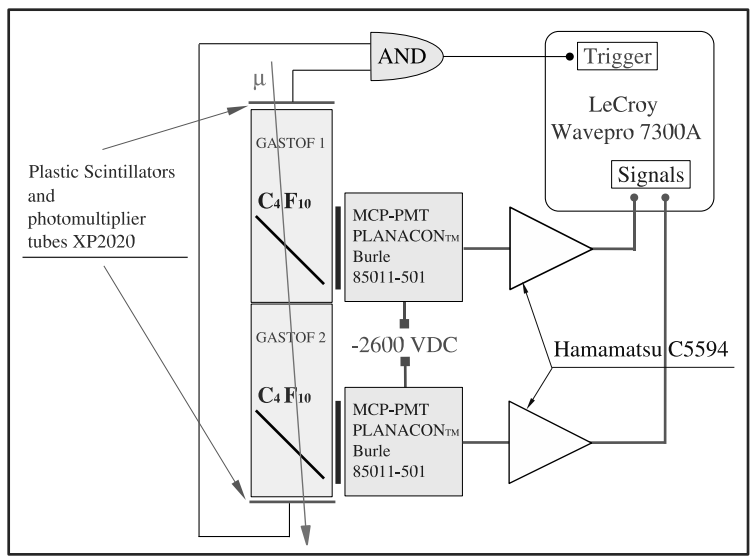,width=100mm}
\end{center}
\caption
{Cosmic ray test stand for the GASTOF prototypes. Plastic scintillators (up and down)
readout by Philips photomultiplier tubes XP2020 are used as a trigger.
The signals from the Burle MCP-PMTs are sent via short SMA cables to very fast Hamamatsu 
C5594 amplifiers. Signals from the GASTOF prototypes as well as from the trigger are read 
using a fast 3GHz LeCroy Wavepro 7300A scope.
}
\label{CosmicStand}
\end{figure}

The GASTOF test stand has been prepared using a simple cosmic ray telescope, as sketched in 
Fig.(\ref{CosmicStand}). Two small plastic scintillator blocs separated vertically and 
readout by Philips PMTs XP2020 are used as a cosmic muon trigger. In each MCP-PMT the 4x4 central group of 
anodes was connected together by short wires of equal lengths, and the rest of anodes was grounded.
The signals from the Burle MCP-PMTs are sent via about 20 cm long SMA cables to very fast Hamamatsu 
C5594 amplifiers. Two GASTOF prototypes placed one after the other and were tested simultaneously.
Two GASTOF signals as well as those from the trigger are read using a fast 3GHz LeCroy Wavepro 7300A 
scope with digital resolution of 50 ps \cite{ref:LeCroy}.

\begin{figure}[!ht]
\begin{center}
\epsfig{file=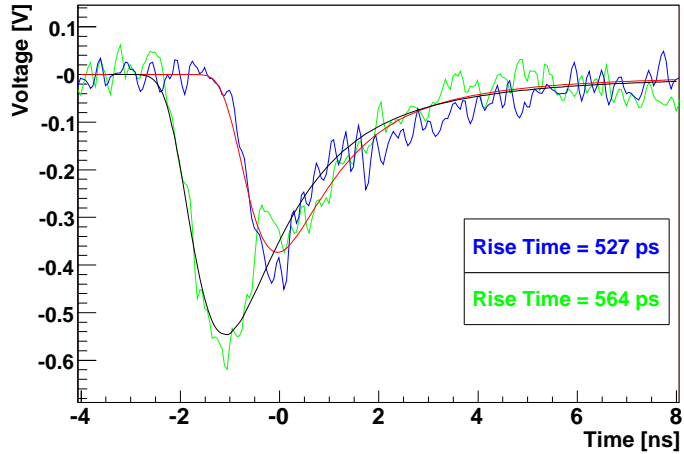,width=100mm}
\end{center}
\caption
{An example of cosmic ray signals from two GASTOF detectors.
The data were taken with the LeCroy Wavepro 7300A scope with digital resolution
of 50 ps. The Landau distribution function was used for fits.
}
\label{result}
\end{figure}

In Fig.(\ref{result}) an example of cosmic ray signals from two GASTOF detectors is shown.
More than hundred such events were collected in a one day run, allowing to make first statistical analysis
of the data. The signals were fitted using the Landau distribution function, and the crucial parameters 
were extracted -- the arrival and rise time of each pulse. The average rise time is similar for both
detectors and is about 600 ps. This is two times worse than the single anode rise time quoted by Burle, 
therefore it is believed to be worsened by the anode grouping. The time difference distribution was also
measured -- it is of gaussian shape with about 100 ps width. Assuming that this width is dominated by the
time resolution of two prototypes, and that they are the same, the upper limit of 70 ps on a single 
detector resolution can be set. This is about two times bigger than the transit time jitter expected for 
the Burle MCP-PMT, but is consistent with the degradation of the Burle rise time, possibly due to the 
anode grouping.

Results from these first tests are encouraging, showing a big potential of GASTOF detectors in domain of 
ultra-precise timing applications. Next steps will include use of the single anode readout or improved
anode grouping scheme, as well as of new Burle MCP-PMTs with even smaller transit time jitter.


\end{document}